\documentclass{PoS}

\title{Study of the VHE diffuse emission in the central 200 pc of our Galaxy with H.E.S.S.}

\ShortTitle{Galactic Center Ridge as seen by H.E.S.S.}

\author{Lemiere A.\\
         APC, In2p3/CNRS, Univeriste Paris 7, Paris, France\\
        E-mail: \email{alemiere@apc.in2p3.fr}}
\author{Terrier R.\\
	        APC, In2p3/CNRS, Univeriste Paris 7, Paris, France\\
                E-mail: \email{terrier@apc.in2p3.fr}}
\author{ Jouvin L.\\
        APC, In2p3/CNRS, Univeriste Paris 7, Paris, France\\
        E-mail: \email{ljouvin@apc.in2p3.fr}}
\author{Marandon V.\\
	MPIK, Heidelberg, Germany\\
        E-mail: \email{marandon@mpi-hd.mpg.de}}
\author{ Khelifi B.\\
        APC, In2p3/CNRS, Univeriste Paris 7, Paris, France\\
      E-mail: \email{khelifi@apc.in2p3.fr}}
\author{for the H.E.S.S. collaboration}

\abstract{The very high energy emission from the  Galactic Center Ridge was revealed by the High Energy Stereoscopic System (H.E.S.S.) in 2006, after subtraction of the point sources HESS J1745-290, possibly associated with Sgr A$^\star$, and HESS J1747$-$281, associated with the composite supernova remnant G0.9$+$0.1. The hard spectrum of the Ridge emission and its spatial correlation with the local gas density suggest that the emission is due to collisions of multi-TeV cosmic rays with the dense clouds of interstellar gas present in this region. The much larger H.E.S.S. dataset (250 hrs) that is now available from this region and the improved analysis method dedicated to the detection of faint emission allow us to reconsider the characterization of this gamma-ray emission in the central 200 pc of our Galaxy through a detailed morphology study. To test the various contributions to the total gamma-ray emission, we use a 2D maximum likelihood approach that allows to constrain a phenomenological model of the signal. We discuss the nature of the various components, and their implication on the cosmic-ray distribution in the central region of our Galaxy. Finally, we will reveal an additional source in this region and will discuss its potential nature.}

\FullConference{The 34th International Cosmic Ray Conference,\\
		30 July- 6 August, 2015\\
		The Hague, The Netherlands}

\begin{document}
\section{Introduction}
The central 200 pc of the Milky-Way are a very rich environment for high energy astrophysics. Besides the closest supermassive black hole (SMBH), Sgr A$^\star$, with its 4 million of solar masses, they contain several tens millions solar masses of molecular gas \cite{Morris96}, hundreds of young and massive stars and their remnants \cite{Mauer07}. Particle acceleration and $\gamma$-ray production are therefore expected to be abundant in this region. 
Observations at Very High Energy (VHE) have shown that the emission from the Galactic Centre (GC) is dominated by the contribution of the point source HESS J1745$-$2901, located at less than 8'' from Sgr A$^\star$ \cite{hess10_GC_position}, and HESS J1747$-$281, associated with the composite supernova remnant G0.9$+$0.1 \cite{hess05_G09}. 
After subtracting these two point sources, a diffuse emission extending over 1$^\circ$ in longitude was discovered by H.E.S.S. with $70$h of data \cite{hess_GC_diffuse}. This emission was found to correlate with dense gas traced by the CS molecule emission \cite{Tsuboi99} indicating that the emission is truly interstellar and due to the collisions of energetic hadrons with the ambient gas. Comparing the $\gamma$-ray flux to the total mass present in the region implied an average density 
of Cosmic-Rays (CR) of energy beyond a few TeV in excess by a factor of 3-9 with respect to the locally measured value. The spectrum of this emission is found to have a power law index of 2.3$\pm$0.1 which is also significantly harder than the locally measured CR flux. All this is a solid proof that the GC region is a site of current or very recent particle acceleration.  A comparison of the $\gamma$-ray profile and the column density of dense gas in longitude revealed an apparent deficit of TeV $\gamma$-rays beyond 1.0$^\circ$$-$1.3$^\circ$ or $130-160$ pc at the GC distance $($the two profiles can be reconciled after multiplying the gas column density by a Gaussian of $\sigma = 0.8^\circ$$)$\cite{hess_GC_diffuse}. This was interpreted as the result of a massive impulsive injection of CRs by a source close to the GC.
Since then several authors have discussed other possible origins for the GC cosmic-ray excess.
In this work, we reconsider the GC ridge emission taking advantage of the much larger H.E.S.S. dataset and the improved analysis methods now avalaible. We use a 2D morphology analysis based on a iterative maximum likelihood method to properly determine the morphology of the ridge emission from $\gamma$-ray images and compare it to maps of gas tracers. In particular, we study emission components that do not follow the general trend given by gas tracers to search for new sources or diffuse excesses of $\gamma$-rays. 
We first present the dataset and the data reduction technique employed, in particular the 2D maximum likelihood fitting technique. We present the various components required to reproduce the data. We report the detection of a new $\gamma$-ray source at a position consistent with the Pulsar Wind Nebula (PWN) candidate G0.13$-$0.13 \cite{Wang03}. We also 
show the existence of an extended central component around Sgr A$^\star$.  
\section{Analysis}
\subsection{H.E.S.S. observations and data analysis}
The data set selected  for  the  analysis includes observations within $5^{\circ}$ of the Galactic Center position carried out from $2004$ to $2012$ with the H.E.S.S. four telescope array.
After standard quality selection and keeping only best quality data ($4$ telescope runs) 
we selected a total of $596$ runs for a total livetime of $259$ hours. The data have a mean zenith angle of $22^{\circ}$.
Data  have  been  analysed with an advanced multivariate analysis procedure developed within the HAP-Fr pipeline, which improves the signal-to-background separation, important  in  searches  for  weak  signals and morphological studies of very 
extended sources \cite{Bech11}.  Hard cuts have been used 
 including a charge threshold of $150$ p.e. per camera image,  
providing an improved angular resolution of $R_{68 \% }=0.077^{\circ}$, which is crucial for this particular analysis, as well as an increased average energy threshold of $~ 350$ GeV. 
To generate $2$D images and background maps, we used an adaptative ring background method \cite{Berge07} with proper exclusion regions, suitable when analysing crowded regions. 
All presented results have been cross$-$checked with an alternative analysis chain 
using independent data calibration, instrument response functions and analysis software designed for the H.E.S.S. Galactic Plane Survey (HGPS) \cite{HGPS}.
In order to be able to compare the total extent of the molecular matter and the $\gamma$-ray emission in the central 200 pc of the Galaxy,  we chose a sufficiently large region that covers the full Central Molecular Zone (CMZ) extension from $1.8^{\circ}$ to $358.5^{\circ}$ in longitude and $\pm 1.5^{\circ}$ in latitude. A  $0.5^{\circ} \times 0.5^{\circ}$ region centered on HESS$~$J1745$-$303 has been removed from the maps in order not to bias the fit with unrelated emission \cite{hess_GC_diffuse}.
For the spectral analysis, the background was estimated using the reflected region method \cite{Berge07}, where the background is derived from circular off-source regions with the same angular size and camera offset as the on-source region. This technique minimizes systematic errors which might be introduced by an incomplete knowledge of the radial acceptance of the camera. The differential VHE $\gamma$-ray spectrum is then fitted with the standard H.E.S.S. software using a forward-folding technique \cite{Piron01}\cite{Jouvin15}.
\subsection{2D Maximum likelihood fit} 
%
%
To test the various contributions to the total $\gamma$-ray emission, we use a 2D maximum likelihood approach. A model of the expected emission is adjusted to the data maximizing its likelihood assuming that counts follow a Poisson distribution in each bin of the $\gamma$-ray map. In each bin $(l,b)$, the observed counts $N_{l,b}$ are assumed to be a realization of a random variable following the Poisson density function with a mean of $\bar{N}_{l,b}$. The latter is the sum of several components:
\begin{itemize}
\item A normalized background event map (the normalized OFF map).
\item A fixed 2D model for the Galactic diffuse emission contribution: unresolved emission along the Galactic plane can produce faint large scale features weighted by the exposure pattern. In order to estimate this contribution we excluded regions with significant emission from the map, and fitted this Galactic component over a large box of $10^{\circ}\times8^{\circ}$, assuming a simple band model (flat along longitude, Gaussian along latitude) convolved by $\gamma$-ray exposure \cite{HGPS}.
\item Sources in the field, mostly HESS$~$J1745$-$290, coincident with SgrA$^\star$, and HESS$~$J1747$-$281, coincident with the composite SNR G0.9$+$0.1, both modeled by a point source for wich position and amplitude are fitted at each step of the fit procedure.
\item The cosmic-ray induced emission is assumed to be the product of the gas tracer density and a 2D symmetrical Gaussian representing the actual CR distribution\footnote{A full 3D model would be better suited to determine the exact distribution, but so far no satisfactory model is available.}. The 2D matter template must be derived from a tracer of the interstellar gas distribution where data is available everywhere along the CMZ. In this study we use  the CS$($1$-$0$)$ line emission \cite{Tsuboi99}. As CS is an optically thin tracer of dense gas, it probes the distribution of dense clouds in the CMZ. 
\item Other components may be added in order to improve the model, depending on the features seen in the 
residual map obtained at each step of the process.
\end{itemize}    
 The H.E.S.S.  instrument response functions are taking into account in the calculation of the model. Indeed, all the model components are convolved with the H.E.S.S. point spread function (PSF), and all are weighted with the $\gamma$-ray acceptance $($produced using the IRFs assuming a spectral index of 2.3$)$. The definition and optimization of the model of the $\gamma$-ray sky describing both the diffuse emission and the sources is done through an iterative n-step process, 
starting with model $0$ including the background map normalized to 1, the diffuse Galatic emission contribution, and the two point sources HESS$~$J1745$-$290 and HESS$~$J1747$-$281, then 
adding at each step of the fit a new component to the model $($models 1,2,3,4 in Table 1$)$ and 
creating a significance map of the residual and a TS map at each step, allowing to search for significant features that could be included in the model.
This iterative fitting process is performed using the $Sherpa$ software package (CIAO v4.5.). The principal model parameters values are given at each step in Table 1 as well as the final model and its corresponding residual maps (in terms of significance level) (see Figure 1).
The significance of sources is measured by the test statistic $\triangle \rm  TS= 2 \rm  log ( \rm  L_{n+1}/ \rm L_{n}$), comparing the likelihood between model n and model $n+1$. 
An associated detection significance can then be calculated for a given source added to the model: the detection threshold was set to 
4$\sigma$.
In order to quantify the goodness-of-fit, we use the Sherpa statistic CSTAT divided by the number of degrees of freedom, that is of the order of 1 for good fits. 
We also compute the statistical significances of the residuals at each position of the map following 
Li$\&$Ma (1983), using the counts map as the number of detected events and the fitted model as the number of expected events (Figure1). The distribution of these residual significances can then be compared with those expected from a large number of Monte Carlo Poissonien simulations following the model (Figure 2). 
\section{Results}
In this section we present our main results from the template-based likelihood fit analysis, and describe the different components as they are added step by step to the model (Table 1). 
First a Large Scale component (LS) modeled as a 2D Gaussian with position fixed on SgrA$^\star$ is used (normalization, $\sigma_{\rm X}$ and $\sigma_{ \rm Y}$ are free). 
Then the CS template (with free amplitude) multiplied by a Gaussian centered on the position (l$=0$,b$=0$) with a free extension $\sigma$ is implemented.
A Central extended Component (CC) defined as a 2D symmetric Gaussian centered on SgrA $^\star$ is added to the model (extension $\sigma$ and amplitude are free). 
 Finally a small 2D Gaussian is included to account for the new source HESS$~$J1746$-$285 (position, amplitude, $\sigma_{\rm X}$ and $\sigma_{ \rm Y}$ are free). 
The full model reproduces the data well with typical residuals that are well distributed and consistent with MC simulations (see Table 1, Figures 1$\&$2). 
The main contribution $($$50 \%$ of the total ridge emission$)$ that is modeled by a matter tracer multiplied by a Gaussian $(\sigma=1^{\circ})$ is not really physically motivated but provides the characteristic scale of the CRs extension. An additional central component with an intrinsic extension $0.1^{\circ}$ and a flux of $32 \%$ of the GC source ($15\%$ of the ridge) is also required to reproduce the data. 
A large scale component extending $\pm30$pc in latitude and $\pm 150$ pc in longitude is also necessary to reproduce the ridge emission, its origin will be discussed in the next section.  
Finally we detect a new source at Galactic position $(l= 0.14^{\circ}\pm0.013^{\circ}, b = -0.114^{\circ}\pm0.02^{\circ} )$, that we name HESS$~$J1746$-$285 (called Arc Source in Table 1). This source is detected at more than 6 $\sigma$ over the CMZ contribution and is compatible with a point-source given the uncertainties. 
A direct output of this analysis is also the production of a significance map of the ridge emission subtracted from the bright point-sources contributions as they have been fitted in the model (Fig.1). 
\begin{table}
\centering
\begin{tabular}{c c c c c c c}      
\hline\hline              
   step       & Model             & Parameter values    &    $\Delta$TS /d.o.f.    & CSTAT   \\
           & components   &                                 &    significance     &     /d.o.f.  \\
\hline\hline
 1    &      LS           & $\sigma_{\rm x} = 0.65 - 0.19^{\circ}+0.21^{\circ}$ , $\sigma_{\rm y} = 0.13^{\circ} \pm0.06^{\circ}$    & 1574/ 3                &  1.05   \\
       &                          &                                                                                                                &  22.9  $\sigma$    &     \\
\hline
 2    &  LS                &  $\sigma_{\rm x} = 0.60^{\circ} -0.16^{\circ} +0.2^{\circ}$ , $\sigma_{\rm y} = 0.19^{\circ} \pm0.05^{\circ}$     & 160/3     &   1.044     \\
     &   CS$\times$Gauss  &    $\sigma  = 1.02 ^{\circ}\pm 0.2^{\circ}$                                                                       &   16.5  $\sigma$      &       \\
\hline
3     & LS                &  $\sigma_{\rm x} = 1.050^{\circ} -0.017^{\circ} +0.02^{\circ}$ , $\sigma_{\rm y} = 0.25^{\circ} \pm0.05^{\circ}$      &   66.6/2  &    1.041    \\
     &   CS$\times$Gauss  &    $\sigma  = 0.98^{\circ} \pm0.19^{\circ}$                                                                        &   8 $\sigma$    &        \\
     &  CC              &     $\sigma  = 0.12^{\circ} \pm0.02^{\circ}$                                                                        &     &        \\
\hline
  4   &  LS                &  $\sigma_{\rm x} = 0.91^{\circ}-0.01^{\circ}+0.02^{\circ}$ , $\sigma_{\rm y} = 0.21^{\circ}\pm0.05^{\circ}$        &   45/5  &     1.039   \\
     &   CS $\times$Gauss &   $\sigma  = 1.10^{\circ}\pm0.20^{\circ}$                                                                            &   6.32 $\sigma$    &        \\
     &   CC             &    $\sigma  = 0.10^{\circ}\pm0.01^{\circ}$                                                                          &     &        \\
     &  ArcSource  &       l $= 0.140^{\circ}\pm0.013^{\circ}$  ,   b $= -0.114^{\circ}\pm0.021^{\circ}2$                                     &     &        \\
    &                            &      $\sigma_{\rm x} = 0.033^{\circ}\pm0.031^{\circ}$ ,   $\sigma_{\rm y} = 0.01^{\circ}\pm0.015^{\circ}$           &     &        \\
\hline\hline
\end{tabular}
\caption{The results of the iterative fit are summarized. Details of the more relevant parameters values are given for 4 steps of the fitting process. The parameters errors $90\%$ confidence intervals (1.64485$\sigma$) are also indicated. The significance of the components detections are given through the  $\Delta$TS /d.o.f. value and converted into equivalent significances. The CSTAT/d.o.f. goodness-of-fit is given as an indication of the fit quality improvement 
at each step of the iterative procedure.} 
\label{table:1}   
\end{table}
\begin{figure}
\begin{center}
\includegraphics[width=1.0\textwidth]{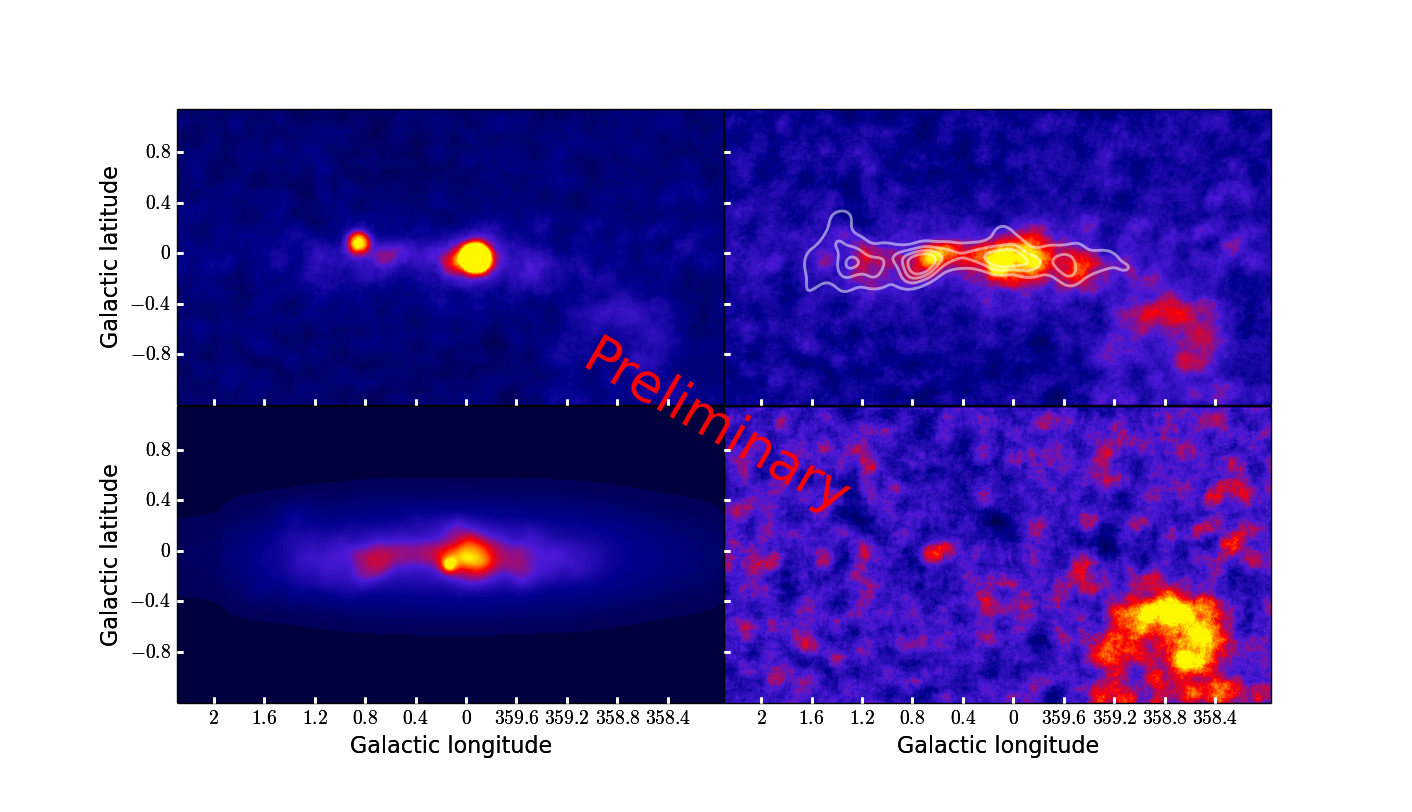}
\caption{\textit{Top Panels: New VHE $\gamma$-ray images of the GC region as seen by H.E.S.S., in Galactic coordinates and smoothed with the H.E.S.S. PSF. Left panel: $\gamma$-ray significance map. Right panel: same map after subtraction of the two point sources G0.9$+$0.1 and HESS J1745$-$290. The white contours indicate the density of molecular gas as traced by the CS emission and smoothed with the H.E.S.S PSF. Bottom left panel:  best model of the GC Ridge VHE emission obtained with this study. Bottom right panel: map of the residual significances obtained with model 4.}} 
\end{center}
\end{figure}
\section{Discussion}
\subsection{HESS$~$J1746$-$285}
Figure 2 shows the confidence-level contours on the best fit position of HESS$~$J1746$-$285 superimposed on a 2$-$10 keV exposure corrected $Chandra$ image.  The source is coincident with an X$-$ray non thermal filamentary structure called G0.13$-$0.11 associated with a point-source, that has been proposed to be a PWN candidate \cite{Wang03}. An examination of the radio images shows some emission but no clear counterpart, although it is clear from the figure that HESS$~$J1746$-$285 and its counterpart G0.13$-$0.11 lie just between the non-thermal filaments of the Radio Arc \cite{Yusef-Zadeh84} (prominents linear filaments perpendicular to the Galactic plane near $\rm l= 0.2^\circ$) and a dense molecular cloud called 
G$0.13-0.13$ \cite{Oka01} believed to be expanding into this Radio Arc.  
The luminosity derived from Chandra observations for G0.13$-$0.11 is L=$3 \times 10^{33} \rm erg/s$, with a spectral photon index of 1.4$-$2.5\cite{Wang03}. 
In such a confined PWN, the X-ray emission results from the synchrotron emission of shocked pulsar wind particles that are pushed downstream into a tail in the direction of the strong and ordered magnetic field. In these conditions, the synchrotron cooling should dominate, making other processes such as inverse
Compton scattering for TeV $\gamma$-ray production much less efficient. 
The intrinsic spectrum of  HESS$~$J1746-285 is estimated to have a flux of $\rm  F(1 \rm TeV)= (1.8\pm 0.33) ×10^{-13}\rm cm^{-2}\rm s^{-1}\rm TeV^{-1}$ and an index of $2.19\pm 0.16$, which gives a luminosity of $ \rm  L=2-3 \times 10^{33}  \rm erg.s^{-1}$ at 8 kpc and a luminosity ratio $L_X/L_{\gamma} \sim 1$, which is in the range of observed PWNe in the Galaxy. However, the origin of the mechanisms yielding the TeV emission is uncertain, in particular given the low density expected in the interstellar bubble of hot shocked gas (so-called Radio Arc Bubble) that fills the area, and the high magnetic field ( $\geq 50 \mu \rm G$). 
Within the error interval, we also found a GeV source detected by the $Fermi-LAT$ at the position (l$=0.14^{\circ}$,b$=-0.11^{\circ}$) identified as 2FGL J1746.6$-$2851c in the second catalog and  1FHL J1746.3$-$2851 in the 1FHL catalog \cite{1FHL}, 
with a large flux above 1 GeV of $6.6 \times 10^{-9} \rm ph.\rm cm^{-2}. \rm s^{-1}$ and a very soft index of $3.2  \pm 0.3$, making the connection with HESS$~$J1746-285 and G0.13$-$0.11 difficult to interpret.
\subsection{Diffuse Emission}
Thanks to the 2D component extraction approach, we confirm that a fraction of GC ridge emission is distributed 
like dense gas tracers. This emission is found to follow closely the CS template up to a projected distance of 
$\sim$ 1.0$^\circ$ or 140 pc. The flux of this component is found to be $\sim$ half of the total GC ridge emission.
Its origin is hadronic CRs interacting  with dense gas. The dip in $\gamma$-ray emission beyond 100$-$150 pc
is likely the result of a combination of decreasing  CR density with distance to the GC and the matter distribution
along the line-of-sight. 
\begin{figure}
\begin{center}
\includegraphics[width=0.48\textwidth]{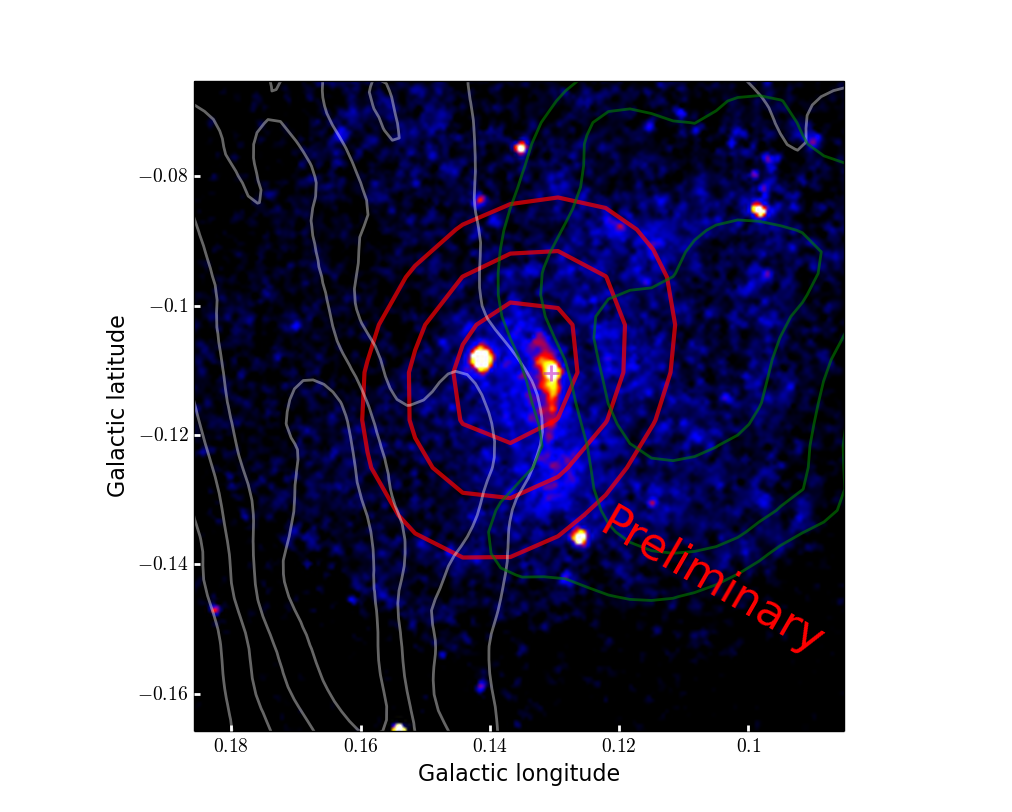}
\includegraphics[width=0.49\textwidth]{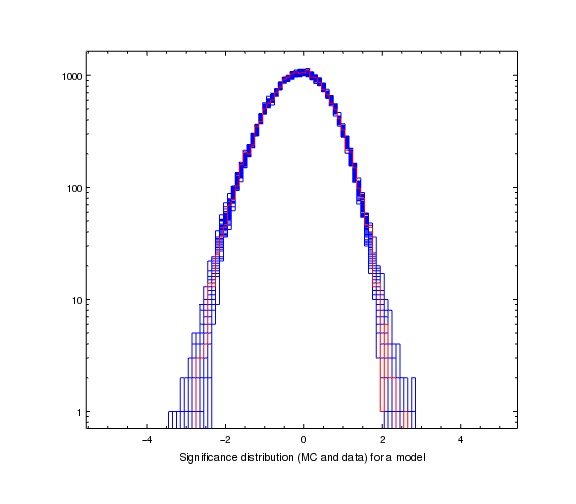}
\caption{\textit{$Left panel$: Chandra 2$-$10 KeV  X$-$ray mosaic image of the field surrounding 
G0.13$-$0.11 using all public data available. The image is exposure-corrected and smoothed with a Gaussian with a width of $\sigma$=2.5'' to highlight 
the filamentary emission. 
Red contours overlaid show the confidence$-$level contours at 68$\%$, 95$\%$, and 99$\%$ on the best fit position of the new detected source HESS$~$J1746$-$285 derived from the TS map of the fitted position (from model 4), white contours show the Radio Arc emission and green contours show the HCN emission of the molecular cloud G0.13$-$0.13. $Right panel:$ Distribution of the significances of the residual map obtained with the data and model4 (in red). In blue we superimpose the significances of the residual map obtained for 100 Poissonian MonteCarlo drawn on the model 4.}}
\end{center}
\end{figure}
In addition, we find that a large scale emission is required to reproduce the observed morphology. This component 
does not correlate with dense gas tracers; in particular, its latitudinal extension is larger. Thus its origin is unclear. A diffuse gas component
 not well accounted for by dense gas tracers such as CS, might be responsible for this component. 
It has been estimated that about 30\% of the molecular gas in the region is found in a diffuse phase with $\sim$ 100 cm$^{-3}$ 
density\cite{Dahmen98}. CR interacting with this H$_2$ phase should contribute to the emission. A number of unresolved sources 
not significant enough to be detected individually probably contribute as well. 
Finally, an interesting feature is found in the central component that the fit requires to reproduce the observed
2D morphology. It is centered on the GC and its radius is about 14 pc. 
It can be the signature of a radial gradient of CRs in the CMZ as the combination of the ridge component with the 
central excess is similar to an integrated 1/r profile \cite{Pevatron}. Such a profile is expected when a stationary source of CRs
is present. This is evidence that a fraction of CRs pervading the CMZ is accelerated at the GC, possibly around 
the SMBH itself. Unresolved sources might also contribute to this central excess, in particular SNRs given the high supernova rate in the CMZ.
Yet, the soft thermal X$-$ray emission that traces SNRs in the CMZ is more extended (up to 0.2$^\circ$-0.3$^\circ)$ around the GC.   
The central 30 pc also contain about 10-15 candidate PWNe.  
If HESS$~$J1746$-$285 is indeed produced by the PWN G0.13$-$0.13, it is to be expected that the population of fainter PWNe clustered 
in the central tens of pc will create a diffuse component in VHE $\gamma$-rays. We note however that a uniform survey of these objects over the CMZ is lacking, and their spatial distribution is therefore rather uncertain.
\section{Conclusion}
We confirm that $\sim$ half of GC ridge emission is distributed 
like dense gas tracers over a projected distance of 140 pc and fades beyond. In addition, we find a large scale emission that does not correlate with dense gas tracers and could be the result of unresolved sources and/or a gas component in a diffuse phase not seen by gas tracers. An additional feature is found in the central $\sim 30$ pc that could be the signature of a radial gradient of CRs in the CMZ that is expected if they are accelerated by the SMBH itself. We finally detected a new source, dubbed HESS$~$J1746$-$285, spatially coincident with the $Fermi$ Arc source and the X$-$ray PWN candidate G0.13$-$0.13.
The question of a physical association should be further studied through by a complete modelling, given the very complex and highly magnetized environment of the object.
\acknowledgments 
The support of the Namibian authorities and of the University of Namibia in facilitating the
construction and operation of H.E.S.S. is gratefully acknowledged, as is the support by the German
Ministry for Education and Research (BMBF), the Max Planck Society, the German Research
Foundation (DFG), the French Ministry for Research, the CNRS-IN2P3 and the Astroparticle Interdisciplinary
Programme of the CNRS, the U.K. Science and Technology Facilities Council (STFC),
the IPNP of the Charles University, the Czech Science Foundation, the Polish Ministry of Science
and Higher Education, the South African Department of Science and Technology and National
Research Foundation, and by the University of Namibia. We appreciate the excellent work of the
technical support staff in Berlin, Durham, Hamburg, Heidelberg, Palaiseau, Paris, Saclay, and in
Namibia in the construction and operation of the equipment.

\end{document}